# The spread of COVID-19 increases with individual mobility and depends on political leaning

*Authors: Chris Parker, Jorge Mejia, and Franco Pestilli*

**Authors' affiliation**

1. Chris Parker, Assistant Professor, Department of IT & Analytics, Kogod School of Business, American University, 4400 Massachusetts Ave NW, Washington, DC 20016, https://orcid.org/0000-0003-1690-770X, URL: https://chrisparker.io, EMAIL: chris.parker@american.edu
2. Jorge Mejia, Assistant Professor, Department of Operations & Decision Technologies, Kelley School of Business, Indiana University, HH4100, 1309 E. 10th St., Bloomington, IN 47405 https://orcid.org/0000-0002-4479-1209 URL: https://go.iu.edu/1T13, EMAIL: jmmejia@iu.edu
3. Franco Pestilli, Associate Professor, Department of Psychology, The University of Texas, Austin,  108 E. Dean Keeton St., Austin, TX 78712 https://orcid.org/0000-0002-2469-0494 URL: https://liberalarts.utexas.edu/psychology/faculty/fp4834, EMAIL: pestilli@utexas.edu

## Abstract

The implementation of social distancing policies is key to reducing the impact of the current COVID-19 pandemic. However, their effectiveness ultimately depends on human behavior. In the United States, compliance with social distancing policies has widely varied thus far during the pandemic. But what drives such variability? Through six open datasets, including actual human mobility, we estimated the association between mobility and the growth rate of COVID-19 cases across 3,107 U.S. counties, generalizing previous reports. In addition, data from the 2016 U.S. presidential election was used to measure how the association between mobility and COVID-19 growth rate differed based on voting patterns. A significant association between political leaning and the COVID-19 growth rate was measured. Our results demonstrate that political orientation may inform models predicting the impact of policies in reducing the spread of COVID-19. [143]

**Significance Statement.**
The right of governments in limiting individual freedom to contain the spread of a pandemic has been contested. Scientists agree that fewer lives are lost when an effective and timely response is implemented. The conversation in the U.S. has argued over the influence of political views on the response to the COVID-19 pandemic. Yet, as of today, evidence lacks as to whether political views might have played a role in influencing human response to the national travel ban order and, in turn, impacted the spread of the pandemic. An association between political leaning and citizens' mobility after the national travel ban is reported. The findings clarify political factors that may have contributed to the spread of the COVID-19 pandemic. [120]

**Acknowledgments.** NSF OAC-1916518, NSF IIS-1912270, NSF IIS-1636893, NSF BCS-1734853, NIH NIBIB 1R01EB029272-01, and a Microsoft Faculty Fellowship to F.P.

## Introduction

In lieu of innovation and pharmaceutical solutions to limit the spread of COVID-19, human behavior-based interventions have been deemed critical to curtailing the spread of the pandemic. Such interventions rely on promoting physical distance between individuals and reducing group gatherings (1, 2). However, adherence to distancing behavior has varied significantly across individuals in the United States. Because of the major divide in the public rhetoric in describing the pandemic, individuals in different states, or with different political beliefs, may experience the risks of the pandemic in strikingly different ways. Here, we tested whether county-level political leaning is associated with social-distancing behavior and the growth rate of COVID-19 cases.

In the U.S., local, state, and federal agencies have imposed orders mandating varying degrees of social distancing (3). For example, these policies have included restrictions in workplace practices, public gatherings, and travel. Recent research has shown that differences in the speed and magnitude at which policies are imposed or relaxed by governments can influence the spread of COVID-19 and impact the economy (4–6) In some states, however, these orders and policies have also been complemented with efforts to communicate the severity of the pandemic and persuade citizens to adapt their behavior.  For instance, NY state officials have used televised briefings, which have been watched by millions, to inform citizens on the dangers of COVID-19 (7), whereas other state officials, such as in FL, have consistently downplayed the severity of the pandemic (8). A more in-depth evaluation of the epidemic process requires a deeper understanding of how government policies, as well as efforts to engage with citizens, may ultimately influence human behavior (9, 10). Indeed these strategies (i.e., mandates, direct communications from officials) are deeply interrelated because the success of the policies in reducing the spread of COVID-19 is likely to depend on the degree to which individuals comply with the mandates and social norms. After all, in the view of many U.S. public officials, it would not be possible to enforce compliance with all social distancing orders.

We investigated whether variables related to the political leaning of individuals influenced the growth rate of the COVID-19 cases. We used open data from several sources, such as the University of Maryland's COVID-19 Impact Analysis Platform (11) and Massachusetts Institute of Technology's Election Data and Science Lab (12)) to investigate how individuals' mobility behavior changed in response to the pandemic and distancing orders implemented by the U.S. government. The UMD data contains a number of county-level metrics. We utilize the cumulative total number of COVID-19 cases to calculate the COVID-19 case growth rate and the social distancing index to construct a measure of individual mobility. The data is at the county-day level, and we utilize data from January 1, 2020, to June 20, 2020. The MIT data contains county-level voting results from the 2016 presidential election. Combining these two datasets allows us to investigate three relationships. First, increased levels of individual mobility are associated with high levels of COVID-19 case growth rates. Second, in early 2020, there was generally no relationship between political leaning and individual mobility, but a strong positive relationship developed in mid-March. Finally, there is an association between political leaning and the differentiation in the COVID-19 growth rates before and after the national travel ban. The results can inform researchers about currently underutilized variables that may improve models of spreading of the pandemic (9).

A substantial amount of scientific evidence has used spreading models (9) to predict the impact of government policies to reduce mobility on the extent and effects of the COVID-19 pandemic (13, 14). For example, using these models, it has been shown that the effective implementation of social distancing policies is associated with reductions in the growth rate of COVID-19 cases, with an average reduction ranging from 3.0% during the early onset days of social distancing measures to 8.6% in the later days (15). Recent findings reported an association between socio-economic factors and compliance to social distancing behaviors (16). Yet, the variability in compliance with social distancing policies across the U.S. and their relation to political facts has not been documented yet.

## Results

**Reduction in individual mobility contributes to controlling the spread of the pandemic**
The U.S. county-level data from the UMC19 dataset was used for all analyses (11). We used linear regression to estimate the COVID-19 case growth rate in each county across the U.S. as a function of individual mobility. In particular, we estimated the following specification:

$$GR_{it} = \alpha_i + \delta_t + b\, IM_{(i,t-14)} + e_{it} \quad \text{Eq. [1]}$$

where $i$ indexes counties and $t$ indexes days. $GR_{it} = ln(cc_{it} + 1) - ln(cc_{it-1} + 1)$ measures the COVID-19 case growth rate as a percentage change in the cumulative number of cases, where $cc_{it}$ is the cumulative number of COVID-19 cases on a given day, and $cc_{it-1}$ is the cumulative number of COVID-19 cases on the previous day. $IM_{(i,t-14)}$ is an individual mobility metric from fourteen days prior to the focal day. To compute individual mobility, we reversed the social distancing index in the UMC19 data: individual mobility = 100 - social distancing index. An individual mobility score of zero essentially means that all the residents in a county are staying home, and no non-residents are entering the county, while an individual mobility score of 100 means that there is no social distancing in the county. $\alpha_i$ is a set of county-level fixed effects that control for any time-invariant characteristics of a county, and $\delta_t$ is a set of date-level fixed effects that control for any characteristics that impact all counties on a given day. The $b$ coefficient is of particular interest as it quantifies the relationship between COVID-19 growth rates on a given day and the 14-day lagged individual mobility across all counties and days, controlling for county and time effects. The term $e_{it}$ represents residual error. The errors were clustered at the state level to allow for arbitrary correlation across counties and days within a state and heteroscedastic error variances across states.

The results show an aggregate association between the COVID-19 case growth rate and 14-day lagged individual mobility of 0.0015 (p-value < 0.001). In other words, for the United States as a whole, a one-point increase in individual mobility (out of 100) is associated with an increased COVID-19 cases growth rate of 0.15%. This result is consistent with findings of a relation between individual mobility and COVID-19 cases growth rate reported by Courtemanche et al. (2020) (15).

Whereas the results based on Eq. [1] demonstrate that, on average, there is a positive association between COVID-19 case growth rates and individual mobility, we also explored whether there are different associations for each county with the following specification:

$$GR_{it} = \alpha_i + \delta_t + b_i IM_{(i,t-14)} + e_{it} \quad \text{Eq. [2]}$$

where all variables are as described in Eq. [1]. The $b_i$ coefficient was of particular interest as it showed the relationship between COVID-19 growth rates and individual mobility for each county individually. The results from Eq. [2] show substantial variability in the association between individual mobility and COVID-19 cases growth rate across the U.S. counties (**Figure 1**). For 66.6% of counties, the association was positive, indicating that increased individual mobility was associated with an increase in the COVID-19 growth rate. We restricted this to only coefficients, which had a p-value below 0.05 and found that 68.8% of those counties had a positive association. The associations varied greatly, with the range from -0.004 to 0.003. Put another way, while in New York County, New York a one-point increase in individual mobility (out of 100) is associated with an increased COVID-19 cases growth rate of 0.3%, in Dawson County, Nebraska that same one-point increase is associated with a decreased COVID-19 cases growth rate of 0.4%.

**Individual mobility and COVID-19 cases growth rate**

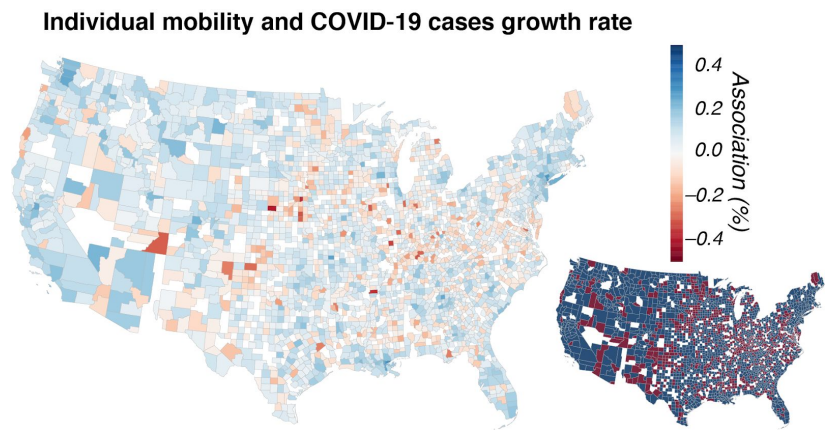

**Figure 1. A positive association between increased individual mobility and COVID-19 cases growth rate.** The colormap of the large U.S. map displays the coefficient $b_i$ from Eq. [2] as a percentage. Coefficients $b_i$ with *p*>0.05 were set

to zero (white) using error clustering at the state level to allow for arbitrary correlations across counties (and days) within each state and heteroscedastic error variances across states (17). The inset map shows in blue the counties with a positive association between individual mobility and COVID-19 cases growth rate, and in red the counties with a negative association. Counties lacking statistical significance are shown in white.

**Individual mobility is associated with political leaning**

After observing the profound variation in the association between individual mobility and COVID-19 cases growth rate, we were interested in understanding possible underlying causes for such variation. We set out to understand whether variations in political leaning across counties could account for part of the variance in the data. This was a plausible and particularly interesting question given the polarized state of American politics and correspondingly divided media reporting on the COVID-19 pandemic. Such a divide may have had an influence on individuals' responses to the national travel ban. In particular, we were interested in measuring whether individual mobility was associated with political leaning. Did individuals with different political leaning in the 2016 elections respond differently to the national travel ban?

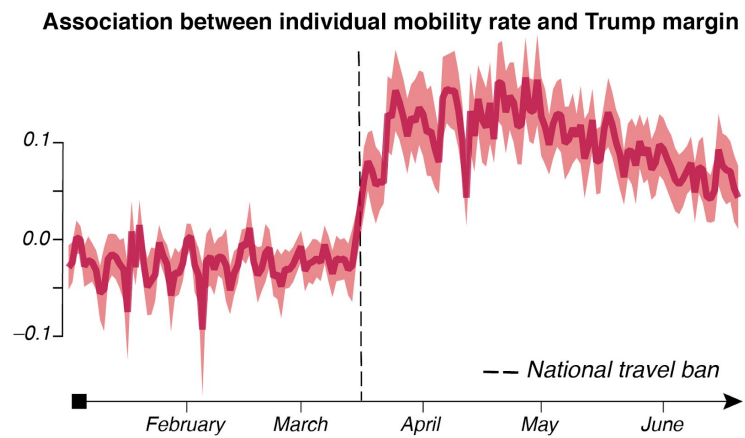

**Association between individual mobility rate and Trump margin**

Figure 2. Change in individual mobility correlation with Trump margin in the 2016 elections. The plot shows parameter $d_t$ of Eq. [3] ±95% confidence intervals based on errors clustered at the state level.

We estimated the association between individual mobility and the 2016 Trump margin (from the MITED dataset) across counties for each day from January 1, 2020 to June 20, 2020 (**Figure 2**). The data shows that before the national travel ban of March 14, 2020, no association was measured between individuals' mobility and the 2016 Trump margin. Critically, after the national travel ban, the association quickly grew positive and remained strong through the end of June 2020. The results show that individuals with different political orientations behaved differently in response to the national travel ban.

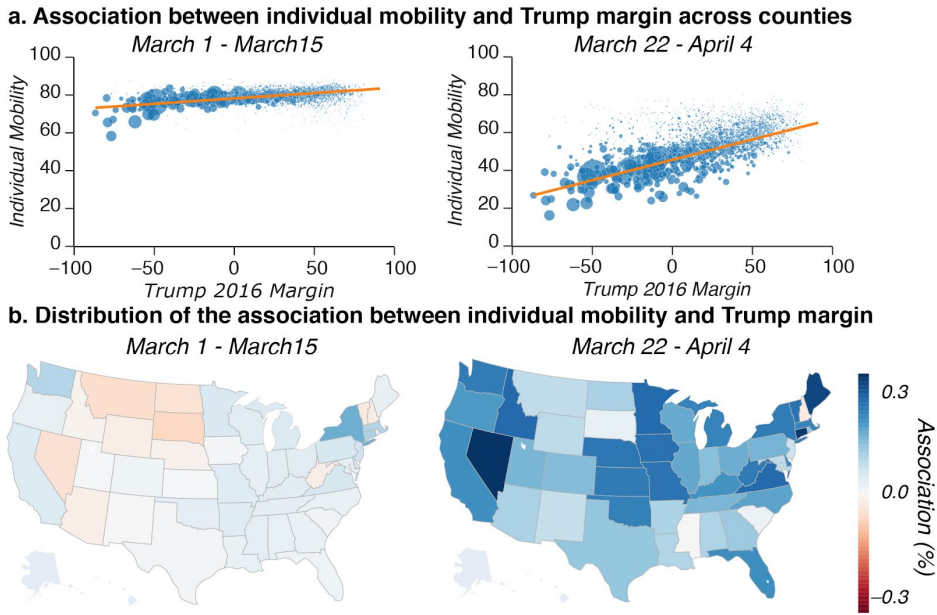

**Figure 3. County and state-level association between individual mobility and 2016 Trump margin before and after the national travel ban. a. County-level individual mobility and Trump margin before and after the travel ban.** We show the trend line showing that the relation between mobility and political leaning goes from zero (baseline; left) to significant after March 14 (right). On this date the national travel ban was issued. The size of each dot represents the population of the county. **b. State-level distribution of the association between individual mobility and Trump margin.** Within a state slope of the orange lines. A darker blue means a higher association. It is darker blue after the travel ban.

We further described how the individual mobility data varies as a function of the 2016 Candidate Trump margin at the county and state level, both before and after the national travel ban (see **Methods**, Eq. [1]). The time-averaged county-level individual mobility data between March 1, 2020 and March 15, 2020 and March 22, 2020 and April 4, 2020 show a not significant and a significant association, respectively (**Figure 3a**, *right- and left-hand panel respectively*). The association is well behaved, and it was found to be largely unrelated (0.057) with Trump's 2016 margin before the national travel ban (*left-hand panel*). In contrast, the association is strong (0.216) after the national travel ban (*right-hand panel*). We further looked at how these associations varied across the United States. **Figure 3b** reports the slope of the regression line measured across each state's counties. Consistent with the previous plot, no clear relationship between individual mobility was found before the national travel ban. In contrast, the state-wide association was strong after the travel ban. Critically, this association varied largely across states with a minimum of -0.029 in New Hampshire and a maximum of 0.345 in Nevada.

**Politics can explain changes in COVID-19 case growth rate**

Having shown a relationship between COVID-19 case growth rates and individual mobility and a relationship between individual mobility and Candidate Trump's 2016 margin, we then explored the association between COVID-19 case growth rates and Candidate Trump's 2016 margin. To do so, we predicted the difference in COVID-19 cases growth rate before and after the national travel ban:

$$\overline{GR_{(i,A)}} - \overline{GR_{(i,B)}} = \alpha + T_i + e_i \quad \text{Eq. [4]},$$

where $i$ indexes a county, $B$ ($A$) indexes the period before (after) the travel ban, and $GR$ is the COVID-19 case growth rate as detailed in Eq. [1], and the bar indicates an average over the days in the before or after period. $T_i$ is the Trump margin in the 2016 election from the MITED dataset. Results show a positive association between Candidate Trump's 2016 margin and the difference in COVID-19 case growth rates before and after the national travel ban (**Figure 4**).

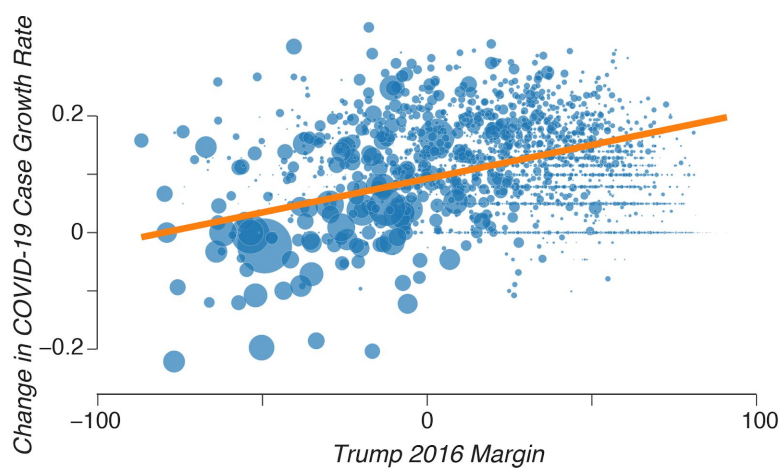

**Figure 4. A positive association between the change in COVID-19 case growth rate and 2016 Trump margin.** The change in COVID-19 case growth rates after the travel ban relative to that county's level before the travel ban is positively associated with Trump's 2016 margin. The size of each dot represents the population of the county.

To further test the robustness of this finding, we performed a variation of the last regression (Eq. [4]) by adding additional control variables (see **Methods**, Eq. [5]). These variables included measures of population size, population density, education, urbanization level, poverty, employment, and income, as well as variables accounting for the five-year range for each county median age. Results show that the association between the change in COVID-19 case growth rate and Candidate Trump's margin was significant even after controlling for all these variables (t-test with 3,830 degrees of freedom; p-value < 0.001).

## Discussion

The unprecedented gravity of the COVID-19 pandemic has mobilized governments for designing a series of ad-hoc policies that can limit citizens' individual mobility and freedom at multiple levels. Many of these policies involve limiting access to basic infrastructure, public offices, schools, and restrict access to travel (3). In the U.S., over 90% of the adult population has had to change their lifestyle as a result of this pandemic (18). Until scientists can produce a vaccine, we are left with behavioral remedies, such as limiting individual mobility, to reduce the impact of COVID-19.

To truly understand how to limit the spread of the disease, it is vital to go beyond government mandates and policies (19) and capture the multifaceted factors that may influence human behavior and the response to policies. One factor of interest is the relationship between political leanings and social distancing behavior. Much like the severity and policies to the pandemic have varied across the U.S., individual compliance with social distancing policies also has also been observed to vary within the population across states and often along political lines. This geographical variation begs the question: are conservatives and liberals in the U.S. complying with the social distancing efforts to the same degree? These efforts include reductions in individual mobility and social activities perceived by many as a reduction to individual freedom. This line of reasoning was perhaps best expressed by the U.S. Surgeon General, the leading spokesperson on matters of public health of the U.S. federal government, "Some feel face coverings infringe on their freedom of choice- but if more wear them, we will have more freedom to go out" (20). Indeed, a recent national poll reported that the percentage of Democrats who wore a mask all the time when leaving home was approximately 65% between May 8, 2020 and June 22, 2020 whereas the percentage of Republicans who wore masks was just 35% (21).

The divide in the current political environment in the United States may suggest that political leaning may play a role in shaping the response to social distancing prescriptions and indirectly affect the spread of COVID-19. First, the U.S. federal government has sent mixed signals throughout the pandemic (22). The administration guidelines have asked Americans to reduce mobility and comply with local laws and wear masks in public spaces to slow the spread of the pandemic and yet, in most cases, President Trump does not wear masks in public (23). Similar mixed signals have been reported by state and local officials (24). Moreover, media coverage on the severity of the COVID-19 pandemic and addressing the need for social distancing has also varied widely along partisan lines. Indeed, U.S. citizens have been starkly divided on the perception of the severity of the pandemic since the beginning of the outbreak. Public discourse from top political leaders, as well as the diversity of opinions expressed by the media, might have played a role in this divide. For example, according to the Pew Research Center, in February 2020, 66% of Democratic-leaning citizens believed the news media coverage of COVID-19 was "largely accurate," while only 31% of Republican-leaning citizens did (25). This, along with comments in social media from prominent figures (i.e., President Trump), may explain why in early March, approximately 40% of Republicans were "not at all concerned" about an outbreak in their communities, whereas less than 5% of Democrats were in that category (26). To make matters worse, U.S. citizens have also been prey to disinformation campaigns regarding COVID-19 from foreign actors, according to the U.S. Department of Defense (27). All these elements taken together make the inclusion of political beliefs in disease growth models relevant, if not necessary.

Recent reports demonstrated that socio-economic factors can play a role in social distancing behaviors with effects on COVID-19 (16). Along the same lines, we find that while limiting individual mobility is associated with reductions in the COVID-19 growth rate, we confirm that people are heterogeneous in the extent to which they adhere to the government policies and social norms, this is associated with varying levels of case growth rates. Furthermore, we demonstrate with county-level data that this association between growth rates and individual mobility is itself correlated with Candidate Trump's 2016 margin of victory. The result: a stronger Trump political leaning in your county is associated with higher COVID-19 growth rates. It is worth stating that due to the nature of the data (secondary), this does not imply a causal relationship.

Understanding how political beliefs may be associated with particular behaviors is crucial for developing and evaluating policies to reduce mobility but also to reopen the economy as it is being attempted in most states. Our findings may be particularly relevant in certain rural regions that supported the President in 2016 and where the first wave of COVID-19 may have been delayed (28). Carefully tracking case growth rates in these areas and pro-actively developing messaging that can influence and motivate human behavior in rural regions may be crucial to reducing the spread of COVID-19 in the coming months.

## Online Methods

### Data sources

*University of Maryland COVID-19 data (UMC19).(11)* fThe data is updated daily and contains data from January 1, 2020, until a few days before the data was downloaded. The data we use for analysis was downloaded on June 24, 2020, and contains data from January 1, 2020, until June 20, 2020. We primarily use the county-level data restricted to only the continental United States (Alaska and Hawaii were excluded from all of the analyses because we encountered problems matching counties to the other datasets). More specifically, the columns used from the UMC19 dataset were: the county name, county- and state-level Federal Information Processing System (FIPS) codes, number of new COVID-19 cases, population, and population density. We adjust the population to be measured as a population in thousands by dividing the population by 1000. We create the individual mobility metric as 100 minus the social distancing index and the lag individual mobility metric as the individual mobility metric in a county but from two weeks (fourteen days) prior. In order to create the growth rate variable, we calculate the cumulative number of cases up to a given day. The case growth rate is then $ln$(cumulative cases today + 1) - $ln$(cumulative cases yesterday + 1). We only use the state-level data to bring in the state name, which is not included in the county-level data. We merge the state- and county-level datasets based on the state FIPS code, which is present in both files.

*Massachusetts Institute of Technology Election Data (MITED).*(12) This data was downloaded on April 22, 2020. We use the candidate, candidate votes, total votes, year, and FIPS columns in this dataset. We exclude counties that do not have a FIPS code since they could not be merged with the UMD dataset. Examples include Uniformed and Overseas Citizens Absentee Voting Act (UOCAVA) in Maine, Rhode Island Federal Precinct, and Connecticut Statewide write-in. The data is limited to only 2016. The percent of votes for each candidate is calculated as 100 times the number of votes for that candidate divided by the total votes for all candidates. The Trump 2016 Margin is then the percent of votes for Trump in a county minus the percent of votes for Clinton in a county. The data is then merged with the UMD data based on the county FIPS code. Oglala Lakota County in South Dakota does not properly merge and is dropped from the analysis.

*Education Data (EdD).* This data comes from the United States Department of Agriculture Economic Research Service Educational attainment for the U.S., States, and counties, 1970-2018 dataset available at https://www.ers.usda.gov/data-products/county-level-data-sets/download-data/ and was downloaded on April 22, 2020. The three measures of education are 1) percent of adults with only a high school diploma from 2014 to 2018, 2) percent of adults completing some college or associate degree from 2014 to 2018, and 3) percent of adults with a bachelor's degree or higher from 2014 to 2018. Excluded is the percent of adults with less than a high school diploma from 2014 to 2018, which serves as the baseline. We also use the county FIPS code for merging this dataset with others.

*Urban/Rural and Poverty Data (PD).* This data comes from the United States Department of Agriculture Economic Research Service Poverty estimates for the U.S., States, and counties, 2018 dataset available at https://www.ers.usda.gov/data-products/county-level-data-sets/download-data/ and was downloaded on April 22, 2020. The two measures of whether a county is urban or rural are 1) the rural-urban continuum code from 2013, and 2) the urban influence code from 2013. The three measures of poverty are 1) percent of people of all ages living in poverty in 2018, 2) percent of people ages zero to seventeen years living in poverty in 2018, and 3) percent of people ages five to seventeen years living in poverty in 2018. We also use the county FIPS code for merging this dataset with others.

*Age Data (AD).* This data was collected from the United States Census Bureau Population Division Annual County Resident Population Estimates by Age, Sex, Race, and Hispanic Origin: April 1, 2010 to July 1, 2019 dataset available at

https://www2.census.gov/programs-surveys/popest/datasets/2010-2019/counties/asrh/cc-est2019-alldata.csv
and was downloaded on July 13, 2020. This dataset includes the number of people in each of eighteen different age groups. The first seventeen age groups are five years wide (zero to four years, five to nine years, etc.) and the final age group is 85 years or older. We use the data to calculate median age dummy variables. For each county, calculate the five-year range in which the median person's age falls. There are ten ranges that have at least one county's median age. We create a set of nine dummy variables that take the value of one if the county's median age is within that five-year range and zero otherwise. The first five-year range is used as the baseline. We also use the county FIPS code for merging this dataset with others.

*Employment Data (EmD).* This data comes from the United States Department of Agriculture Economic Research Service Unemployment and median household income for the U.S., States, and counties, 2000-18 dataset available at https://www.ers.usda.gov/data-products/county-level-data-sets/download-data/ and was downloaded on April 22, 2020. The measure of employment is the unemployment rate from 2018. The two measures of income are 1) the median household income from 2018, and 2) the median household income as a percent of the state total in 2018. We also use the county FIPS code for merging this dataset with others.

*Plot.ly county geojson (PCG).* The PCG is a set of JSON files openly provided by Plot.ly©. The data allows the mapping of numerical values onto the U.S. map by county or state. This data was used only for visualization purposes (**Figures 1** and **3**). The data is available at
https://raw.githubusercontent.com/plotly/datasets/master/geojson-counties-fips.json.

### Data modeling approach

#### Variation in Association between Individual Mobility and COVID-19 Case Growth Rate Across Counties

We used regression to map the COVID-19 cases growth rate of disease spread in each county across the U.S. In particular, we estimated the following specification:

$$GR_{it} = \alpha_i + \delta_t + b\,IM_{(i,t-14)} + e \quad \text{Eq. [1]}$$

where $i$ indexes a county and $t$ references a day. $GR$ is $ln(cc_{it} + 1) - ln(cc_{it-1} + 1)$, measuring the COVID-19 growth rate as a percentage change in the cumulative number of cases. $cc_{it}$ is the cumulative number of COVID-19 cases on a given day, and $cc_{it-1}$ is the cumulative number of COVID-19 cases on the day before the focal day. $IM_{(i,t-14)}$ is the individual mobility metric from the UMC19 dataset described above from fourteen days before the focal day. $\alpha_i$ is a set of county-level fixed effects that control for any time-invariant characteristics of a county, $\delta_t$ is a set of date-level fixed effects that control for any characteristics that impact all counties on a given day. The $b$ coefficient shows the relationship between the COVID-19 case growth rate and 14-day lagged individual mobility on average across all counties and days. The errors are clustered at the state level to allow for arbitrary correlation across counties and days within a state and heteroscedastic error variances across states.

We then modified Eq. [1] to allow for different coefficients for every county:

$$GR_{it} = \alpha_i + \delta_t + b_i IM_{(i,t-14)} + e \quad \text{Eq. [2]}$$

Our primary interest lies in the $b_i$ coefficients that show the association between the 14-day lagged individual mobility and COVID-19 growth rates for each county $i$.

#### Association Between Individual Mobility and Trump 2016 Margin Over Time

We performed a second linear regression to estimate the association between individual mobility, using U.S. county as fixed effects ($\alpha_i$), the date was also used as fixed effects ($\delta_t$), and date fixed effects interacted with the Trump 2016 margin:

$$IM_{(i,t)} = \alpha_i + \delta_t + d_t \delta_t\, T + e \quad \text{Eq. [2]}$$

where $i$ indexes a county and $t$ references a day, $T_i$ is the Trump margin in the 2016 election from the MITED dataset, and all over variables are the same as in Eq. [1]. Errors are clustered at the state level to allow for arbitrary state-level heteroskedasticity and correlated errors within states across all counties and all dates. The coefficients of the date fixed effect interacted with the Trump 2016 margin are then plotted along with their 95% confidence intervals.

**Association Between Individual Mobility and Trump 2016 Margin Immediately Before and After the Travel Ban**

**Figure 3** was created by running the same regression multiple times. **Figure 3a** uses all county-days for Eq. [3]. **Figure 3b** runs Eq. [3] forty-eight times, one time per state (excluding D.C. because it only has one county as well as Alaska and Hawaii because of lack of data). In each of these runs, the data is limited to only the county-days in that state.

**Figure 3a** was created by limiting the time range of the data. Before is defined as March 1, 2020 to March 15, 2020 and after is defined as March 22, 2020, to April 4, 2020. We then calculate the average individual mobility during that time period for each state. A weighted least squares regression is performed with the dependent variable the average individual mobility, independent variable the Trump 2016 margin, and population for weights:

$$\overline{IM_{(i,B|A)}} = \alpha + T_i + e \text{ Eq. [3]}$$

where all variables are as previously defined. The regression was run once for the before period ($B$) and once for the after period ($A$). The bar indicates an average over the days in the before or after period. The scatter points are sized by the population and the line is the weighted least squares regression line.

Finally, we repeated the above process individually for each state, i.e., using only the county-days for each state. The slope of the line in the weighted least squares regression was then used to color the state in the filled maps in Figure 3b.

**Modeling the Association between the Change in COVID-19 Case Growth Rate Across Counties and Trump 2016 Margin**

The line in **Figure 4** was created from a weighted least squares regression:

$$\overline{GR_{(i,A)}} - \overline{GR_{(i,B)}} = \alpha + T_i + e_i \text{ Eq. [4]}$$

where the dependent variable was the change in COVID-19 case growth rate, the bar indicates an average over the days in the before or after period, the independent variable is the county-level Trump 2016 margin, and weights are based on the population in the county.

We then modified Eq. [4] to include multiple control variables from four complementary datasets:

$$\overline{GR_{(i,A)}} - \overline{GR_{(i,B)}} = \alpha + T_i + Controls_i + e_i \text{ Eq. [5]}$$

where all variables (except $Controls_i$) are as before and the weights are based on the population in the county. The control variables used were: population (UMC19), population density (UMC19), percent of adults with only a high school diploma from 2014 to 2018 (EdD), percent of adults completing some college or associate degree from 2014 to 2018 (EdD), percent of adults with a bachelor's degree or higher from 2014 to 2018 (EdD), the rural-urban continuum code from 2013 (PD), the urban influence code from 2013 (PD), percent of people of all ages living in poverty in 2018 (PD), percent of people aged zero to seventeen years living in poverty in 2018 (PD), percent of people aged five to seventeen years living in poverty in 2018 (PD), nine dummy variables that take the value of one if the county's median age is within that five-year range and zero otherwise (AD), the unemployment rate from 2018 (EmD), the median household income from 2018 (EmD), and the median household income as a percent of the state total in 2018 (EmD).

**Availability of data and reproducibility of code**

All the data used and generated for this paper is publicly available, the sources are reported above. All the data extracted from the original sources and the code used to organize and analyze the data is publicly available as a Jupyter Notebook at <PDF VERSION INCLUDED IN SUBMISSION; URL PROVIDED WITH PUBLICATION>.